\documentclass{elsart}
\usepackage{amsfonts}
\usepackage{amsmath}
\usepackage{amssymb}
\usepackage{graphicx,color}
\usepackage{bm}

\begin{document}

\begin{frontmatter}%

\title{Thermodynamics of Multiple Two-body Systems with Long-range Correlation}%

\author{Luan Cheng\corauthref{cor}}
\corauth[cor]{Corresponding author},Huiqiang Ding, Furong Yan, Weining Zhang%

\address{Institute of Theoretical Physics, Dalian University
of Technology, Dalian 116024, China}%

\ead{luancheng@dlut.edu.cn}%

\author{Enke Wang\corauthref{cor}}%

\address{Institute of Particle
Physics, Central China Normal
University, Wuhan 430079, China}%

\ead{wangek@mail.ccnu.edu.cn}%

\begin{abstract}
We aim to study thermodynamics of multiple two-body systems with
long-range correlation using non-extensive statistics. Long-range
correlation will cause multiple systems in anomalous diffusion. We consider
the influence of long-range correlation as a background noise effect on a two-body system.
 We solve probability and entropy equations of a two-body system to obtain the temperature and distance
 dependence of the non-extensive parameter. The result shows the long-range correlation changes the system's entropy and energy. The more strongly
 is the system bounded, the less its energy is affected by the long-range correlation. Moreover, the anomalous diffusion approaches Brown
motion with increasing temperature. This will help to understand how
nonlinear field affects thermodynamics of a system.
\end{abstract}%


\end{frontmatter}%

\section{Introduction}

Two-body system research is an interesting issue in many areas of
physics. In material physics, the critical temperature of two-electron
system-- cooper pair is used to determine the physical property of
superconducting material. In high-energy physics, the thermodynamical
character of quark-antiquark system--mesons is studied to analyze the physics
in heavy-ion collisions. So thermodynamical research of two-body
systems is significant in many areas of physics.

The two-body system is not isolated. Normally, research objects are composed of
large amount of two-body systems, such as the superconducting material composed
of many cooper pairs, quark-gluon plasma containing many mesons. The electric force
between electrons is a long-range interaction, so that the
electrons in the cooper pair will interact with other electrons in other
cooper pairs. The two-body systems are not free and independent. Instead, they
are long-range correlated. The diffusion of independent two-body systems is
Brown motion. The diffusion of long-range correlated systems is anomalous diffusion\cite{Konkonen}.
 Anomalous diffusion was found in many systems including ultra-cold atoms\cite{Sagi}, Telomeres in
the nucleus of cells\cite{Bronstein}, single particle movements in
cytoplasm\cite{Regner}, worm-like micellar solutions\cite{Jeon}.
Anomalous diffusion was also found in other biological systems, including
heartbeat intervals and in DNA sequences\cite{Buldyrev}. Moreover, anomalous
diffusion environment is found to have significant effect on a system. The
purpose of this paper is to study the thermodynamics of the two-body
system with long-range correlation and analyze the thermodynamical effect.

Long-range interactions will induce system's non-extensiveness\cite{Tsallis}.
Long-range microscopic interactions exhibit singularities at the origin in
Boltzmann-Gibbs(B-G) statistics\cite{Tsallis}. This made B-G statistics an
inefficient theory to describe anomalous thermostatistical behaviour. Then,
 which method can describe long-range interactions
issues a challenge to us. Fractal theory exhibits a repeating self-similar
pattern at every scale. It is also known as expanding symmetry or evolving
symmetry. Non-extensive statistics is a statistical theory by
the light of fractal idea which aims to describe non-extensiveness. In
non-extensive statistics, the parameter $q$ decribes the non-extensiveness
of the system. In the limit of $q\rightarrow 1$ non-extensive statistics
comes to B-G statistics. So the entropic index $q$ in $S_{q}$ can be regarded
as physical effect on a standard B-G system which causes the system's
non-extensiveness and non-additive entropy. Here, we will use non-extensive
statistics to study how anomalous diffusion, which is caused by long-range
correlation, influence the thermodynamics of
the two-body system.

This paper is organized as the following. Section 2 constitutes an
introduction to the non-extensive statistical mechanics theory. Section 3 is
dedicated to solve the probability and entropy equations to obtain the
entropic index $q$, $q^{\prime }$ and analyse long-range correlation effect
on the system energy. In section 4, we will use electrical dipole system as
a case study and analyze the numerical result. Finally, we will come to a
conclusion and discuss the thermal environment effect on the system.

\section{Formalism in non-extensive statistics}

Non-extensive statistical mechanics has been used to describe phenomena in
many physical systems: dusty plasmas\cite{Liu}, trapped ions\cite{DeVoe},
spin-glasses\cite{Pickup}, anomalous diffusion\cite{Huang}, material
physics, high-energy physics\cite{Adare}. Its advantage over B-G statistics
is rather than exhibiting singularities, non-extensive statistics can solve the problem
in long-range interaction and anomalous diffusion.

Non-extensive statistical mechanics is based on the generalized functional
form of the entropy\cite{Tsallis}(in natural unit with $k=\hbar =1$)
\begin{equation}
S_{q}=\frac{1-\sum_{i=1}^{\Gamma}p_{i}^{q}}{q-1}\left( \sum_{i=1}^{\Gamma}p_{i}=1;q\in
R\right) \text{,}  \label{q-entropy}
\end{equation}%
where $\Gamma$ is the number of microscopic states. Here the parameter $q$
describes the non-extensiveness of the system.

In classical version, the non-extensive entropy can be expressed as
\begin{equation}
S_{q}=\frac{1-\int dx[p(x)]^{q}}{q-1}(\int dxp(x)=1).  \label{q-entropy2}
\end{equation}
The B-G entropy
can be obtained in the limit of $q\rightarrow 1$.

\begin{figure}
\centering
\includegraphics[width=12cm]{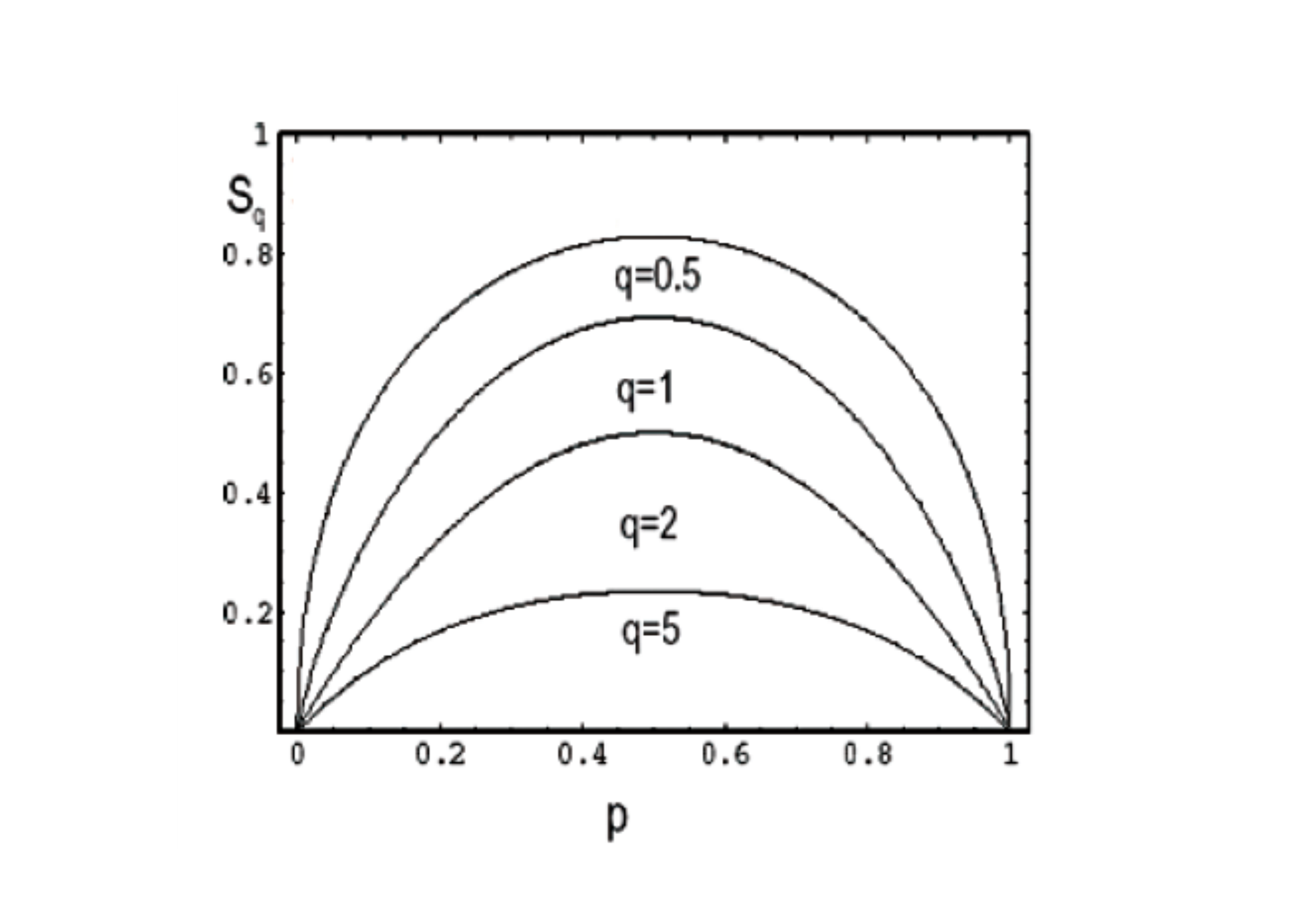}
\caption{The p-dependence of the $\Gamma = 2$ entropy for typical values of $q$.}
\label{fig:sqp}
\end{figure}


Shown in Fig.\ref{fig:sqp} is the non-extensive entropy as a function of
probability $p$ at fixed states number $\Gamma=2$ for typical values of $q$\cite%
{Tsallis}. We can find that the non-extensive entropy increases with
decreasing $q$ at fixed $p$. In other words, $S_q$ is larger than
B-G entropy when $q<1$, less than B-G entropy when $%
q>1$.

It can be straightforwardly verified that for system $A_{1}$ and $A_{2}$ if
the joint probability satisfies $%
p_{ij}^{A_{1}+A_{2}}=p_{i}^{A_{1}}p_{j}^{A_{2}}$, then%
\begin{equation}
S_{q}(A_{1}+A_{2})=S_{q}(A_{1})+S_{q}(A_{2})+(1-q)S_{q}(A_{1})S_{q}(A_{2}).
\label{non-add entropy}
\end{equation}
Due to this property, for $q\neq 1$, $S_{q}$ is said to be
nonadditive.

In classical version, the mean value of a variable (referred to as
the q-mean value) in non-extensive statistical mechanics is:

\begin{equation}
\langle x\rangle _{q}\equiv \int_{0}^{\infty }dxxP(x),  \label{mean value}
\end{equation}

where $P(x)$ is the escort distribution, which is defined as an arbitrary,
possibly fractal, probability distribution. In thermostatistics of
multifractals, the probability distribution $p(x)$ is sought on the basis of
incomplete knowledge, an occurrent event will induce a set of further
probability distribution \cite{Beck}

\bigskip
\begin{equation}
P(x)\equiv \frac{\lbrack p(x)]^{q}}{\int_{0}^{\infty }dx^{\prime
}[p(x^{\prime })]^{q}}.  \label{q-probability}
\end{equation}

We can immediately verify that $P(x)$ is also normalized, i.e.,

\begin{equation}
\int_{0}^{\infty }dxP(x)=1.  \label{normalization}
\end{equation}




It can be understood in this manner that under some physical effects, the
probability, entropy and energy of the system changes from $p(x)$, $S_{B-G%
\text{, }}$and $\langle E\rangle $ with $q=1$ to $P(x)$, $S_{q}$, and $\langle E\rangle
_{q}$ with $q\neq 1$, respectively.

If the system is in equilibrium, with principle of maximum entropy, the
Fermi-Dirac and Bose-Einstein(escort) mean occupation number distributions
could be generalizable as follows\cite{Beck2}

\begin{equation}
n_{i}=\frac{1}{[1+(q-1)\beta (\varepsilon _{i}-\mu )]^{\frac{q}{q-1}}\pm 1},
\label{ni}
\end{equation}

where $i$ corresponds for microscopic state, $\beta $ and $\mu $ are
effective inverse temperature and chemical potential respectively, and $\pm $
correspond to fermions and bosons respectively. In the favor of the
distribution function, it has an impressively good fitting of high
temperature experimental data obtained in electron-positron collisions \cite%
{Bediaga}.

\section{Thermodynamics of two-body system with long-range correlation}

Here the research object is supposed to consist of a large amount of two-body
systems. The interaction between the two bodies inside a two-body system is a
long-range interaction. This will induce one body not only interacts with the other body
inside a two-body system, but also interacts with others in other systems. So
the systems are long-range correlated. The long-range correlation among different systems
will cause anomalous diffusion\cite{Konkonen}. The long-range interaction will make other systems perform work
on a system, change its energy and entropy.

If a two-body system is isolated and free, the probability of this system can be written as
\begin{equation}
P=\frac{e^{-\beta \varepsilon }}{\frac{1}{(2\pi )^{6}}\int e^{-\beta
\varepsilon }d^{3}p_{1}d^{3}p_{2}d^{3}r_{1}d^{3}r_{2}},  \label{B free probability}
\end{equation}%
where $p_{1}$, $p_{2}$, $r_{1}$, $r_{2}$ are momentum and position of the two bodies.
 The energy of the system can be
written as $\varepsilon=\frac{p_{1}^{2}}{2m}+\frac{p_{2}^{2}}{2m}%
+V_{pot}$, where $V_{pot}=V_{pot}(r)$ is the long range interaction
potential between the two bodies inside the bound system, $r$ is the distance between the
two bodies.

Meanwhile, since this system is isolated, according to equal probabilities of the
fundamental hypothesis of equilibrium statistical mechanics, the probability in any
microscopic state can be expressed as
\begin{equation}
P=\frac{1}{\Gamma},  \label{PB2}
\end{equation}%
where $\Gamma$ is the total number of microscopic states of the two-body system.

The mean occupation number distribution of this free two-body system as
an explicit function is
\begin{equation}
n=\frac{1}{e^{\beta \varepsilon}-1},  \label{nBi}
\end{equation}
where $\varepsilon$ is the system's energy.

However, this two-body system is not free, it long-range interacts with other two-body systems.
Then other systems will collide with it,
perform work on it and finally reaches equilibrium. The long-range correlation among the systems will induce
anomalous diffusion\cite{Konkonen}. The long-range correlation will influence the
system and induce probability changes to $P^{\prime }$, energy changes to $\varepsilon ^{\prime }$,
 distribution changes to $n^{\prime }$ and its entropy changes
to $S^{\prime }$.

The energy content of a system consists of the heat which is put into it and
the sum of the work performed on it
\begin{equation}
d\overline{\varepsilon}=dQ+dW.  \label{energy change1}
\end{equation}

The variation of the system's energy $d\overline{\varepsilon}=d(\sum\limits_{i}P_{i}\varepsilon _{i})$ in Eq. (\ref{energy change1})
 can also be considered as a result of the variation of energy level and the variation of the corresponding possibility.
\begin{equation}
d\overline{\varepsilon}=\sum\limits_{i}\varepsilon
_{i}dP_{i}+\sum\limits_{i}P_{i}d\varepsilon _{i}.  \label{energy change2}
\end{equation}

The quantity of heat transferred is given by $dQ=\sum\limits_{i}\varepsilon
_{i}dP_{i}$\cite{schwabl}. A transfer of heat gives rise to a
redistribution of the occupation probabilities. Heating increases the
populations of the states at higher energies.

Energy change by an input of work is
\begin{equation}
dW=Ydy=dy\sum\limits_{i}\frac{\partial \varepsilon _{i}}{\partial y}%
P_{i}=\sum\limits_{i}P_{i}d\varepsilon _{i}  \label{work input}
\end{equation}
where $Y$ is the generalized force, $y$ is the external parameter. This
means the work input produces a change in the energy eigenvalues but doesn't
change the corresponding probability. So the probability $P_{i}$, the
number of microscopic states $\Gamma$ and mean occupation number $n_{i}$
remains the same. Long-range correlation can be considered as outside particles
performing work on the system, so that it does not change the probability
and mean occupation number distribution of the system, but the energy
eigenvalue is changed to $\varepsilon _{i}^{\prime }$. So the probability and
mean occupation number distribution of the two-body system with long-range correlation are\cite%
{Tsallis}
\begin{equation}
P^{\prime }=\frac{(\frac{1}{\Gamma})^{q^{\prime }}}{\sum%
\limits_{i=1}^{\Gamma}(\frac{1}{\Gamma})^{q^{\prime }}}=\frac{1}{\Gamma}=P=\frac{%
e^{-\beta \varepsilon }}{\frac{1}{(2\pi )^{6}}\int e^{-\beta \varepsilon
}d^{3}p_{1}d^{3}p_{2}d^{3}r_{1}d^{3}r_{2} };  \label{PBq}
\end{equation}

\begin{equation}
n^{\prime }=\frac{1}{[1+(q^{\prime }-1)\beta \varepsilon ^{\prime }]
^{\frac{q^{\prime }}{q^{\prime }-1}}-1}=n=\frac{1}{%
e^{\beta \varepsilon }-1},  \label{qdistribution}
\end{equation}
where $q^{\prime }$ is the non-extensive parameter of the two-body system in anomalous diffusion, its departure from $q=1$ reflects
the anomalous diffusion effect on the two-body system. If $q^{\prime }=1$, this system is in Brown motion.

Using Eq.(\ref{q-entropy}), the entropy of the two-body system with long-range correlation
is
\begin{equation}
S^{\prime }=\frac{1-\sum_{i=1}^{\Gamma}{P^{\prime }}^{q^{\prime }}}{q^{\prime
}-1}.  \label{B entropy1}
\end{equation}

We expand the function ${P^{\prime }}^{q^{\prime }}$ around the initial free state $%
q^{\prime }=1$ and terminate the expansion after the quadratic term if $%
q^{\prime }$ is around 1, we obtain
\begin{align}
{P^{\prime }}^{q^{\prime }}& ={P^{\prime }}^{q^{\prime }}|_{q^{\prime }=1}+\frac{\partial
{P^{\prime }}^{q^{\prime }}}{\partial q^{\prime }}|_{q^{\prime }=1}(q^{\prime }-1)+%
\frac{1}{2}\frac{\partial ^{2}{P^{\prime }}^{q^{\prime }}}{\partial q^{\prime 2}}%
|_{q^{\prime }=1}(q^{\prime }-1)^{2}+\cdots   \notag \\
& =P+\ln P\cdot P(q^{\prime }-1)+\frac{(\ln P)^{2}\cdot
P(q^{\prime }-1)^{2}}{2}+\cdots.  \label{PBiq expansion}
\end{align}

Then we can rewrite
the entropy of the two-body system with long-range correlation as
\begin{equation}
S^{\prime }=-\ln P-\frac{1}{2}\ln ^{2}P\cdot (q^{\prime }-1)
\label{SBq}
\end{equation}

So that,
\begin{equation}
q^{\prime }=\frac{-\ln P-S^{\prime}}{\frac{1}{2}\ln ^{2}P}+1.
\label{q'-value}
\end{equation}

Next, we come to analyse the physics inside the two-body system.
If the two bodies(we name them $a_{1}$ and $a_{2}$) are free, the two bodies
 should obey B-G statistics. The probability of body $a_{1}$ is,
\begin{equation}
P_{a_{1}}=\frac{e^{-\beta \varepsilon _{1}}}{\frac{1}{(2\pi )^{3}}\int \int
e^{-\beta \varepsilon _{1}}d^{3}p_{1}d^{3}r_{1}}.  \label{free probability}
\end{equation}
where $\varepsilon _{1}$ is body $a_{1}$'s energy.

For non-relativistic particles, we have $\varepsilon _{1}=p_{1}^{2}/2m$.
Then we can get
\begin{equation}
P_{a_{1}}=\frac{8\pi ^{3}e^{-\beta p_{1}^{2}/2m}}{V(\frac{2\pi m}{\beta }%
)^{3/2}},  \label{free probability2}
\end{equation}
where $V$ is the system's volume, $m$ is the mass of the particle, and $\beta $
is the inverse temperature of the system $1/T$.

Body $a_{1}$ interacts with $a_{2}$ and is affected by long-range correlations from
other systems. These two physical aspects could be
considered as an occurrent event, that will induce the $a_{1}$'s probability change to
\begin{equation}
P_{a_{1q}}=\frac{[P_{a_{1}}(p_{1},r_{1})]^{q}}{\frac{1}{(2\pi )^{3}}\int
\int [P_{a_{1}}(p_{1},r_{1})]^{q}d^{3}p_{1}d^{3}r_{1}}=\frac{8\pi
^{3}e^{-\beta qp_{1}^{2}/2m}}{V(\frac{2\pi m}{\beta q})^{3/2}}
\label{q-probability 2}
\end{equation}%
according to Eq.(\ref{q-probability}) in thermostatistics of multifractals.
Then with Eq.(\ref{q-entropy2}) the changed entropy $S_{a_{1q}}$ can also
be obtained.

The calculation of body $a_{2}$ should obey the same rule as $a_{1}$.

The probability of the two-body system with long-range correlation $P^{\prime}$ is the joint probability of body $a_{1}$ and $a_{2}$, so
\begin{equation}
P^{\prime}=P=P_{a_{1q}}\cdot P_{a_{2q}}.  \label{A-probability}
\end{equation}

The entropy of the two-body system is the sum of the two bodies' entropy and the non-extensive part,
according to Eq.(\ref{non-add entropy}),
\begin{equation}
S^{\prime}=S_{a_{1q}}+S_{a_{2q}}+(1-q)S_{a_{1q}}S_{a_{2q}}.  \label{A-entropy}
\end{equation}

For convenience, we consider that the bodies inside the two-body bound system are static($p_{1}=p_{2}=0$). Solving the equation (\ref{A-probability}),
 $q$ can be obtained,
\begin{equation}
q=\left( \frac{Ve^{-\beta V_{pot}}}{\int e^{-\beta V_{pot}}d^{3}r}\right)
^{1/3}.  \label{q-value}
\end{equation}
With the obtained $q$ in Eq. (\ref{q-value}), one can get body $a_1$ and $a_2$'s entropy and
the entropy of the two-body system with long-range correlation.  Substituting the entropy the two-body system into the Eq. (\ref{q'-value}) gives
the value of $q^{\prime }$. With the obtained $q^{\prime }$, from Eq. (\ref{qdistribution}), the energy
of two-particle system is changed to
\begin{align}
\varepsilon _{q^{\prime }}& =\frac{1}{(q^{\prime }-1)\beta }\big{(}e^{\frac{%
q^{\prime }-1}{q^{\prime }}\beta \varepsilon }-1\big{)}  \notag \\
& \approx \sum\limits_{n=1}^{\infty }[(q^{\prime }-1)\beta ]^{n-1}(\frac{%
\varepsilon }{q^{\prime }})^{n},  \label{energyq'}
\end{align}%
due to long-range correlation among the two-body systems.

\section{Numerical Results}

We take electrical dipole bound system with the point charge $Q=20e$ and
mass $M=1MeV$ as a case study. The
interaction potential between the electrical dipole is $V_{pot}=-\alpha
_{s}/r$ with $\alpha _{s}=1/137$ in natural unit. Because of long-range electromagnetic force,
 other electrical dipole bound systems will have weak interactions
on the dipole bound system and cause anomalous diffusion.

\begin{figure}[tbp]
\centering
\includegraphics[width=13.5cm]{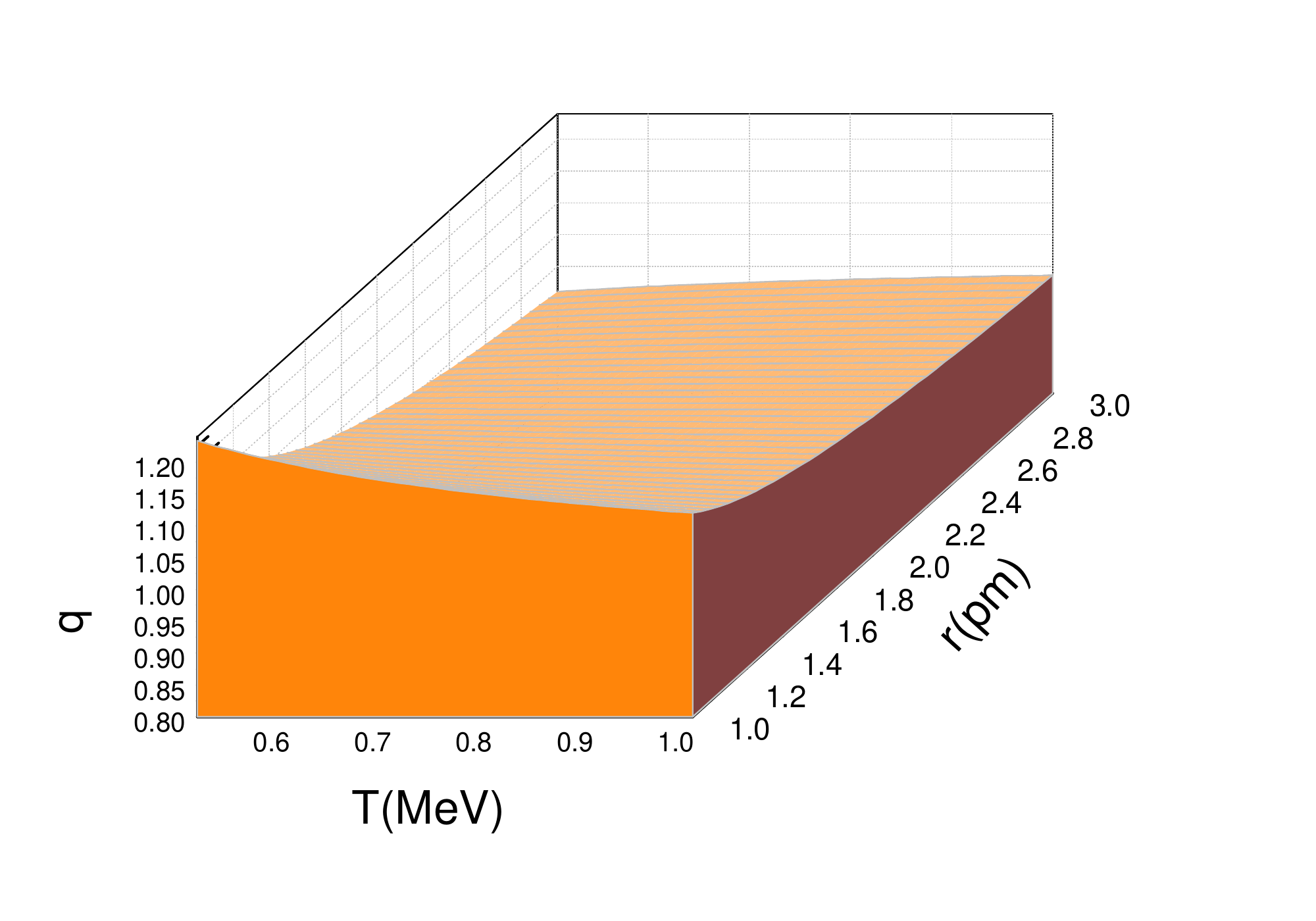}
\caption{The Temperature and the two-body distance dependence
of the non-extensive parameter $q$ of the body $a_1$ in the two-body system.}
\label{fig:qTr3}
\end{figure}

The non-extensive parameter $q$ describes both the attraction and
environmental effect on the body inside the two-body system. We obtained the temperature
 and the two body distance dependence of the non-extensive parameter $q$ of
  the body $a_1$ in the bound system, which is shown in Fig. \ref{fig:qTr3}.
Normally, binding decreases a body's entropy but long-range correlation outside the system will increase it.
Here, our result is the competition of the two effects. Mostly the interaction inside the system affects the body in the system
more than the environmental noise outside. So it is found that here mostly
attraction induces $q>1$. Considering Eq. (\ref{q-entropy}) and Fig. \ref%
{fig:sqp}, the non-extensive entropy decreases with
increasing $q$, so that it shows the bound state decreases the body's entropy if compared to the free state $q=1$.
 However, when the distance is
large and the interaction between the bodies is weak, the environmental noise outside affects more. This induces that $q$
is a bit less than unit and the body $a_1$'s entropy is a little larger than that when $q=1$.
Moreover, the departure
of non-extensive parameter $q$ from $q=1$(B-G statistics) decreases when
increasing the temperature.

\begin{figure}[tbp]
\centering
\includegraphics[width=13.5cm]{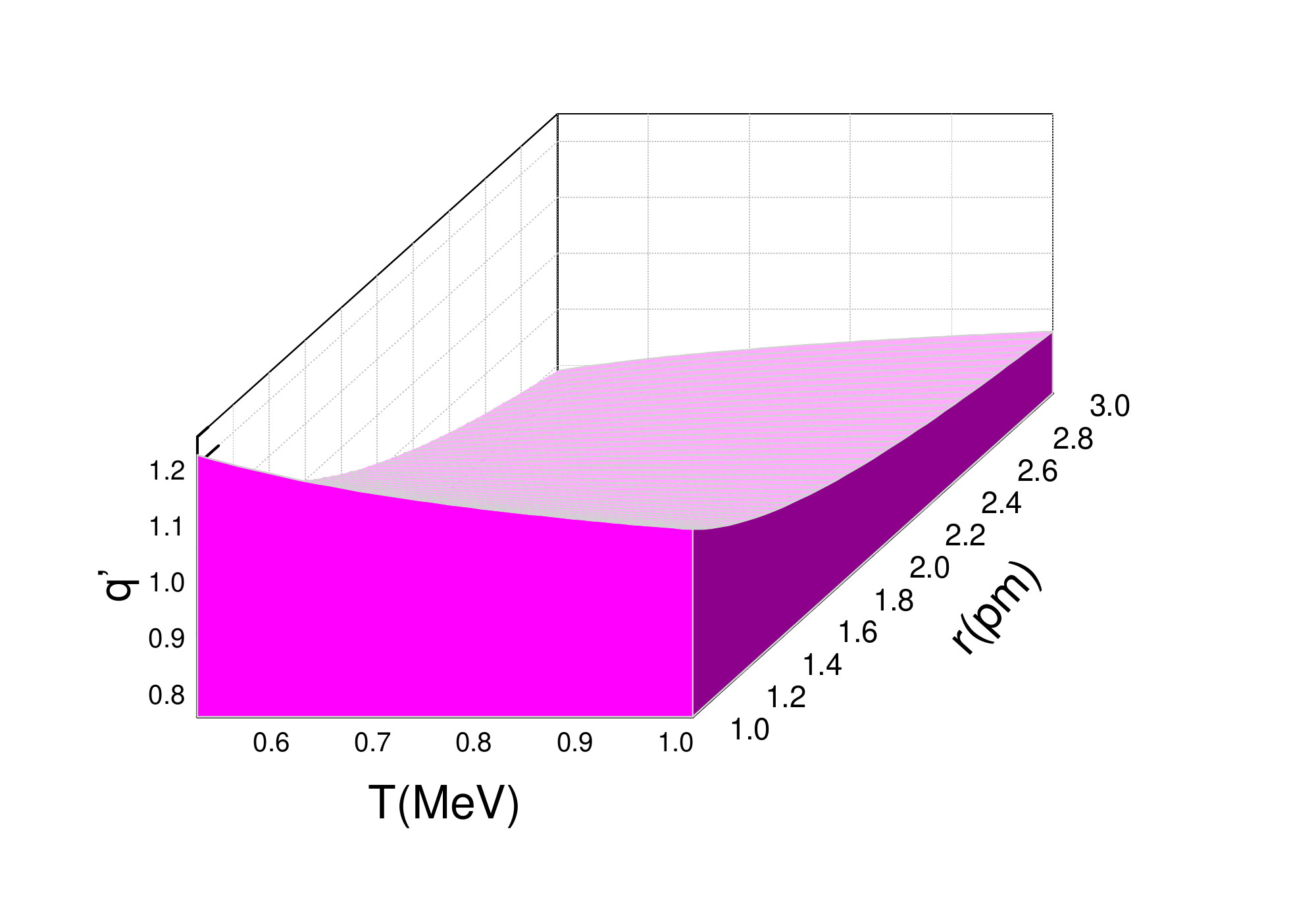}
\caption{The Temperature and the two-body distance dependence
of the non-extensive parameter $q^{\prime }$ of the two-body system with long-range correlation.}
\label{fig:qpTr}
\end{figure}

The non-extensive parameter $q^{\prime }$ describes the
environmental noise effect(long-range correlation effect with other systems) on the two-body system.
 Its departure from unit comes from long-range interaction by other similar systems and work performed on it. We neglect the
approximation in Eq. (\ref{PBiq expansion}) and solve Eq. (\ref{B entropy1}), Eq. (\ref{A-entropy}) numerically,
 then we obtain the temperature and two-body distance
 dependence of $q^{\prime }$, which is shown
in Fig. \ref{fig:qpTr}. It indicates that the interaction inside
the system affect the environment. When the distance between the two
bodies is small and the binding energy is large, it is found $q^{\prime
}>1$. This means this environment is a supper-diffusion environment\cite%
{Tsallis2}. When the distance becomes larger and the binding energy is
smaller, there is strongly overlapping among the similar two-body systems, $q^{\prime }$ turns to be less than unit, which means the
environment is a sub-diffusion one. Secondly, although $q^{\prime }$ is
different at different circumstances, we find that its departure from $q=1$%
(B-G statistics) decreases with the increase of the temperature. That indicates
with increasing temperature, thermal motion of the multiple two-body systems rises while
the influence of the long-range correlation declines, so that the anomalous diffusion
approaches Brown motion.

\begin{figure}[tbp]
\centering
\includegraphics[width=13.5cm]{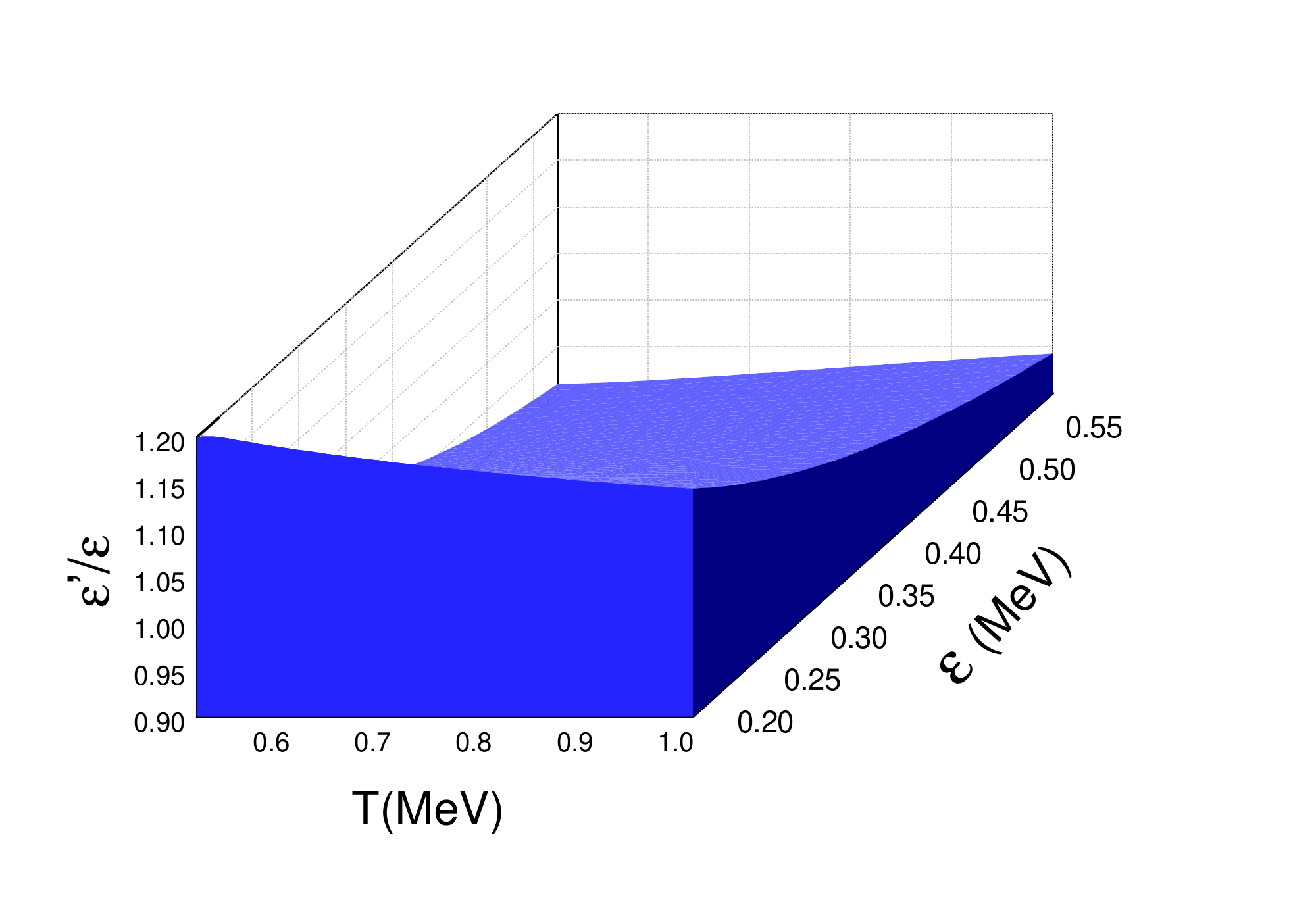}
\caption{The ratio of the system energy in the anomalous diffusion to the
isolated system energy $\protect\varepsilon^{\prime }/\protect\varepsilon$
as a function of temperature $T$ and the isolated energy $\protect%
\varepsilon $.}
\label{fig:epTe}
\end{figure}

Fig. \ref{fig:epTe} illustrates the ratio of the system energy in the
anomalous diffusion to the isolated system energy $\varepsilon^{\prime
}/\varepsilon$ as a function of temperature $T$ and the isolated energy $%
\varepsilon$. It is found that the ratio decreases with the increase of
absorption between the two particles inside the two-body system. This indicates that
the more strongly the system is bounded, the less it is affected by the
anomalous diffusion environment. In order to analyse the validness of our result.
We compare it with BCS-BEC crossover picture, which also describes the two-body
state with an absorption between the two bodies. It is found that the result is consistent with BCS-BEC crossover%
\cite{Levin}. In the BCS phase, the binding inside the cooper pair is weak,
the pair size is large, so that there is strongly overlapping among the cooper pairs, the
environment affects the cooper pair a lot. In the BEC phase, the binding is
strong, the pair size is small, so that preformed pair nearly compose ideal
gas, the environment affect the pair very little. We also find that the
ratio's departure from unit decreases with the increase of the temperature, this
again indicates that the anomalous diffusion approaches Brown motion with
increasing temperature.

\section{Conclusion}

We use non-extensive statistical mechanics to study the thermodynamics of
 multiple two-body systems. We solve probability and entropy equations and obtain the
non-extensive parameter. We also calculate the ratio of the system energy in
the anomalous diffusion to the isolated system energy $\varepsilon ^{\prime
}/\varepsilon $ as a function of temperature $T$ and the isolated energy $%
\varepsilon $. We take an electrical dipole bound system with long-range correlation with the point
charge $Q=20e$, mass $M=1MeV$ and the interaction potential $V_{pot}=-\alpha
_{s}/r$  as a
case study. We consider the non-extensive parameter $q$ of body $a_1$ in the
two-body system. The result is the competition of two effects, absorption between $a_1$, $a_2$
and long-range correlation effect outside system, which is considered as environmental noise .
Mostly the binding inside the system affects the body more than
the environmental noise outside so that it decreases the body's entropy if compared to the free state.
 However, when the distance is
large and the interaction is weak, the environmental noise outside affects more.
This induces the particle's entropy is a little larger than free state. Moreover, the departure of $q$
from $q=1$(B-G statistics) decreases when increasing the temperature.The non-extensive parameter $q^{\prime }$ describes the
environmental noise effect(collision effect from other systems which induces anomalous diffusion) on the system. Both the interaction inside the
system and environmental temperature affect $q^{\prime }$. When the system's
binding energy is large and two-body distance is small, $q^{\prime }$ is larger than unit, the environment
is a supper-diffusion environment. In the opposite condition, there is strongly overlapping among the two-body systems,
 $q^{\prime }<1$ and the environment is a sub-diffusion one. Although $q^{\prime }$ is
different at different circumstances, its departure from $q=1$(B-G
statistics) decreases with the increase of temperature. That indicates
with increasing temperature, thermal motion of the multiple two-body systems rises while
the influence of the long-range correlation declines, so that the anomalous diffusion
approaches Brown motion. We also find that the ratio of the system energy with long-range correlation
to the isolated system energy $\varepsilon ^{\prime }/\varepsilon $
decreases with the increase of absorption between the two bodies inside the two-body
system. The more strongly the system is bounded, the less it is affected
by the long-range correlation outside the two-body system. The ratio's departure from unit also
decreases with increasing temperature, this again indicates that the
anomalous diffusion approaches Brown motion with the increase of temperature.
The result is consistent with BCS-BEC crossover picture. This will help to
understand how nonlinear field affect the system energy.

\section{Acknowledgments}

This work was supported by the National Natural Science Foundation of China
(Grant Nos. 11205024, 11221504,11435004,11275037 and 10825523), the Major State Basic Research
Development Program in China (Grant No. 2014CB845404) and the Ministry of
Education of China (the Doctoral Grant No. 20120041120043).


\begin{thebibliography}{99}
\bibitem{Konkonen} J. Konkonen and E. Karjalainen, J. Phys. A 21, 22 (1988)

\bibitem{Sagi} Y. Sagi, M. Brook, I. Almog, N. Davidson, Phys. Rev. Lett.
108, 093002 (2012)

\bibitem{Bronstein} I. Bronstein, Y. Israel, E. Kepten, S. Mai, Y. Shav-Tal,
E. Barkai, Y. Garini, Phys. Rev. Lett. 103, 018102 (2009).

\bibitem{Regner} B, Regner, D. Vu\v{c}ini\'{c}, C, Domnisoru, T. M. Bartol,
M. W. Hetzer, D. M. Tartakovsky, T. J. Sejnowski, Biophy. J. 104(8),1652
(2013).

\bibitem{Jeon} J. Jeon, N. Leijnse, L. Oddershede, R. Metzler, New J. Phys.
15 (4): 045011 (2013).

\bibitem{Buldyrev}  Bunde, Armin, Havlin, Shlomo. Fractals in Science. Springer.
pp. 49--89. ISBN 3-540-56220-6.

\bibitem{Tsallis} C. Tsallis, Introduction to Nonextensive Statistical
Mechanics---Approaching a Complex World (Springer, New York, 2009);
 U. Tirnakli, G. F. J. Ananos, C. Tsallis, Phys. Lett. A 289, 51 (2001).

\bibitem{Liu} B. Liu and J. Goree, Phys. Rev. Lett. 100, 055003 (2008).

\bibitem{DeVoe} R. G. DeVoe, Phys. Rev. Lett. 102, 063001 (2009)

\bibitem{Pickup} R. M. Pickup, R. Cywinski, C. Pappas, B. Farago, and P.
Fouquet, Phys. Rev. Lett. 102, 097202 (2009)

\bibitem{Huang} Z. Huang, G. Su, A. El Kaabouchi, Q. A. Wang, and J. Chen,
J. Stat. Mech. 2010, L05001 (2010)

\bibitem{Adare} A. Adare et al., Phys. Rev. D 83, 052004 (2011).

\bibitem{Beck} C. Beck and F. Schlogl, Thermodynamics of Chaotic Systems
(Cambridge University Press, Cambridge, 1993).

\bibitem{Beck2} C. Beck, Physica A 286, 164 (2000).

\bibitem{Bediaga} I. Bediaga, E. M. F. Curado, and J. Miranda, Physica A
286, 156 (2000).

\bibitem{schwabl} F. Schwabl, Statistical Mechanics(Springer-Verlag Berlin
Heidelberg 2006).

\bibitem{Tsallis2} C. Tsallis, D. Bukman, Phys. Rev. E 54, 3 (1996).

\bibitem{Xu} S. Gupta, X. Luo, B. Mohanty, H. Ritter, N. Xu, Sicence 332,
1525 (2011).

\bibitem{Biro} T.S. Biro, G. Purcsel, and K. Urmossy, Eur. Phys. J. A 40,
325 (2009).

\bibitem{Levin} Q.Chen, J. Stajic, S. Tan, K. Levin, Physics Reports 412, 1
(2005).
\end{thebibliography}
\end{document}